\documentclass[a4paper,11pt]{article}
\usepackage{jinstpub} % for details on the use of the package, please see the JINST-author-manual
\usepackage{lineno}
\usepackage{romannum}
\usepackage{subcaption}
\title{\boldmath NIOS II Soft-Core Processor and Ethernet Controller Solution for RPC-DAQ in INO ICAL}

\author[a,1]{Yuvaraj Elangovan,\note{Corresponding author.}}
\author[a]{Mandar Saraf,}
\author[a]{B. Satyanarayana,} 
\author[a]{S.S. Upadhya,} 
\author[a]{Nagaraj Panyam,} 
\author[a]{Ravindra Shinde,} 
\author[a]{Gobinda Majumder,} 
\author[a]{D. Sil,} 
\author[a]{Pathaleswar,} 
\author[a]{S. Thoi Thoi,} 
\author[a]{and K.C. Ravindran} 
\affiliation[a]{Tata Institute of Fundamental Research, Mumbai, India}

\emailAdd{yue8@pitt.edu}

\abstract{
This paper introduces a high-performance Soft-Core Processor based data acquisition system  designed for handling Resistive Plate Chambers (RPCs). The DAQ consist of FPGA-based hardware equipped with Soft-Core Processor and embedded hardwired Ethernet controllers named RPC-DAQ, offering a versatile and fast network-enabled data acquisition solution. A soft processor, NIOS, is instantiated within an Intel Cyclone \Romannum{4} FPGA, overseeing control, communication, and data transfer with remote processing units. These integrated RPC-DAQ units, in substantial numbers, connect to a limited set of high-end processing units via LAN switches. This paper provides a detailed account of the software implementation scheme for the NIOS processor in the RPC-DAQ system. A remarkable 28,800 RPC-DAQ units will be deployed in proximity to the RPCs, serving the proposed INO-ICAL experiment in Theni-Madurai, Tamil Nadu. The network-enabled RPC-DAQ units controlled by the soft processor offloads FPGA tasks including event data acquisition, periodic health monitoring of RPCs, command interfaces, high voltage control, and data transfer to back-end data concentrators. Communication and data transfer are executed efficiently via TCP and UDP protocols over a 100 Mbps Ethernet interface. This system provides innovative solutions to improve data acquisition and control in large-scale scientific experiments.}


\begin{document}
\maketitle
\flushbottom

\section{INO-ICAL Overview}
\label{sec:overview}
The INO-ICAL (India-based Neutrino Observatory - Iron Calorimeter) experiment is designed to study the open question of neutrino mass hierarchy using atmospheric neutrinos by detecting muons generated from neutrino interactions with iron plates. This experiment features a 50kton electromagnet consisting of iron plates stacked in a cuboidal geometric configuration, with 28,800 Resistive Plate Chambers (RPCs) positioned between them to detect the charged trajectories passing through the detector. RPCs are well-suited for this purpose, offering exceptional time resolution and reliability in tracking charged particles. Each RPC is paired with an FPGA-based RPC-DAQ (Data Acquisition Module) which is responsible for digitizing signals, including both position and timing data, enabling precise and efficient event processing. An RPC, along with its front-end electronics, is considered as a single detector unit. Each detector unit is equipped with an Ethernet interface through the RPC-DAQ and assigned a dedicated private IP address. These units can be controlled by groups of central servers using standard TCP/IP protocols, enabling seamless communication and management. The R\&D phase is completed, and we are now running a small prototype to validate the full chain of electronics along with all mechanical design. As soon as the clearance (environment etc.) from the government is obtained, this system will be commissioned in a few years’ time scale.

\section{Introduction}
\label{sec:intro}
\begin{figure}[htbp]
\centering
\includegraphics[width=.7\textwidth]{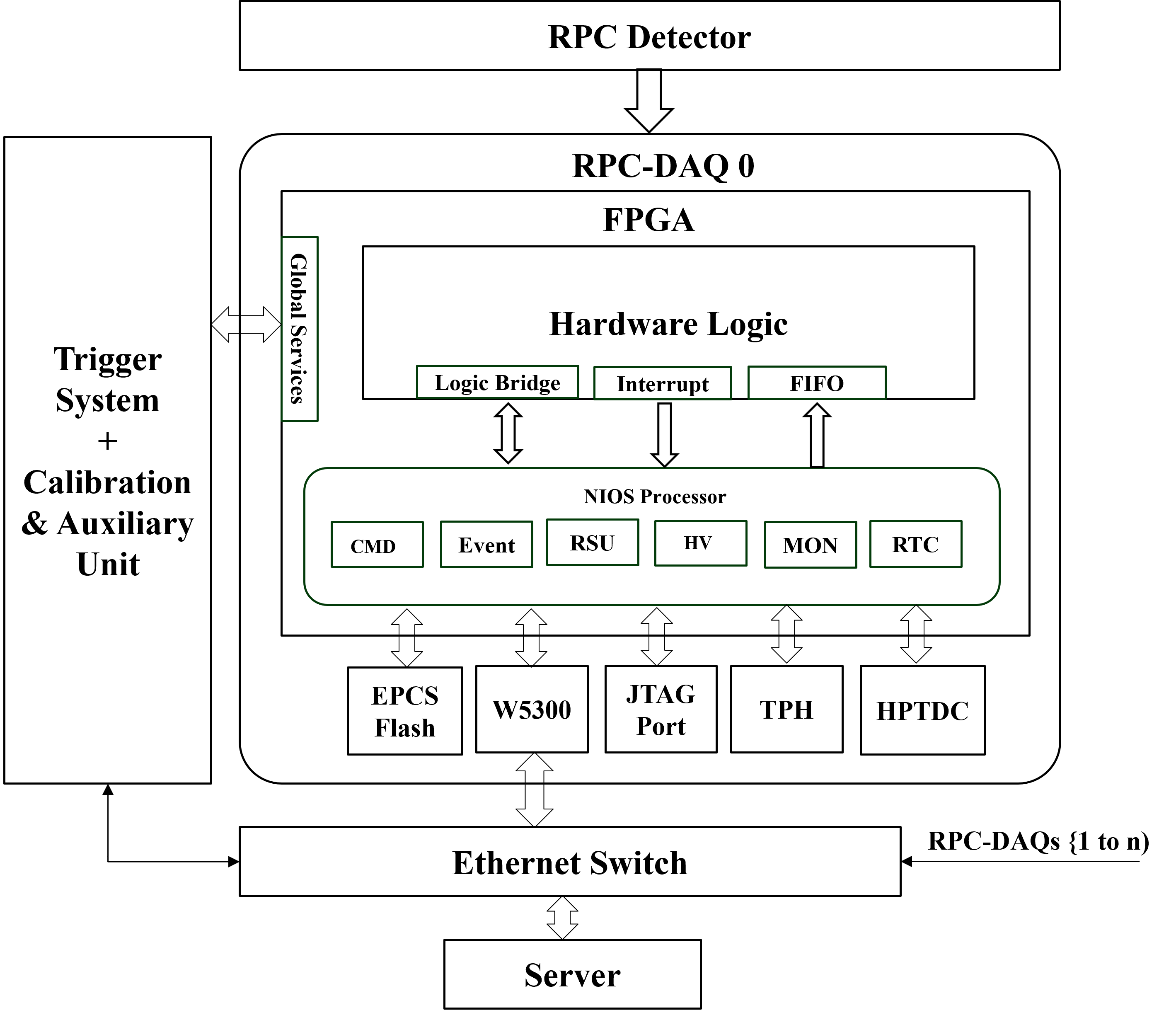}
\caption{NIOS Processor Overview.\label{fig:1}}
\end{figure}

The RPCs in INO ICAL, measuring $2\, \text{m} \times 2\, \text{m}$, features 59-63 strip sensor signals in both the X and Y directions, providing precise XY position information and Time of Flight (TOF) data for particle interactions. Figure~\ref{fig:1} illustrates the comprehensive functionality of the RPC-DAQ module~\cite{a}. A Soft-Core processor NIOS~\cite{b} is instantiated in an Intel Cyclone \Romannum{4} FPGA~\cite{c}, acting as an intermediary between the FPGA hardware logic and the Ethernet controller in RPC-DAQ. NIOS soft-core processors are highly customizable. One can configure the processor's word length (e.g., 16-bit or 32-bit), clock frequency, memory architecture, peripherals and other features which help to minimize the functionality by hardware implementation. Compared to System on Chips (SoCs), NIOS soft core processor uses FPGA resources such as logic elements and block RAM, thus allowing to optimize the usage of the available hardware. This also enables fine-tuning of the design and to meet specific size, performance and power requirements. When a global trigger arrives the NIOS processor reads "XY" strip and timing information from the hardware logic and frames an event packet with a global timestamp. This timestamp is later used by the data servers in event building at the Back-end. The Ethernet controller manages data buffering, transfer, and communication with the back-end servers.

The major responsibilities of the processor are event data recording, monitoring data recording, execution of back-end commands and  remote firmware upgrade (RSU). These tasks are executed through interrupt-driven software routines. When a global trigger occurs the event Interrupt Service Routine (ISR) is invoked. This ISR acquires event data of variable size from the hardware and stores it into the event FIFO. Once the data in the Event FIFO surpasses an optimal packet size, it is pushed to the Ethernet controller's buffer for background transfer to the back-end server by the processor.

A preprogrammed timer in the FPGA hardware logic generates a monitoring interrupt. The Monitoring ISR  collects essential RPC health parameters such as the noise rate of pickup signals, High voltage currents and pre-trigger rates. These parameters are pushed into respective socket FIFOs within the Ethernet controller. Back-end commands received by the Ethernet controller invoke the Command ISR, ensuring the execution of commands with optional acknowledgments. The event ISR, monitoring ISR and command ISR are assigned priorities in descending order to optimize system functionality. The hardwired Ethernet controller Wiznet W5300 uses 6 out of 8 sockets shown in Table~\ref{tab:1} for the communication and data transfer to back-end and each socket is configured with fixed buffer size. 

\begin{table}[htbp]
\centering
\caption{Socket Configuration of the Ethernet Controller.\label{tab:1}}
\smallskip
\begin{tabular}{c|l|l}
\hline
Socket No. &Protocol &Usage \\
\hline
0 & UDP Multicast & Command interface\\
1 & TCP Client & Event Data Transfer\\
2 & TCP Client & Mon Data Transfer\\
3 & TCP Server & Remote System Upgrade\\
4 & UDP Unicast & Command Interface\\
5 & UDP Client & RTC Data Transfer\\
\hline
\end{tabular}
\end{table}
The RPC-DAQ in the INO-ICAL experiment must handle a minimum trigger rate of 2 kHz and transmit event packets containing position and timing information at a throughput of up to 2 Mbps, with an average event size of 120 bytes. Additionally, it is crucial to simultaneously handle the slow control of various elements and ancillary units of the detector as well as monitoring of its performance parameters. 

\section{Command Processing in NIOS}
\label{sec:proto}
\begin{figure}[htbp]
\centering
\begin{subfigure}[c]{0.45\textwidth}
\includegraphics[width=\linewidth]{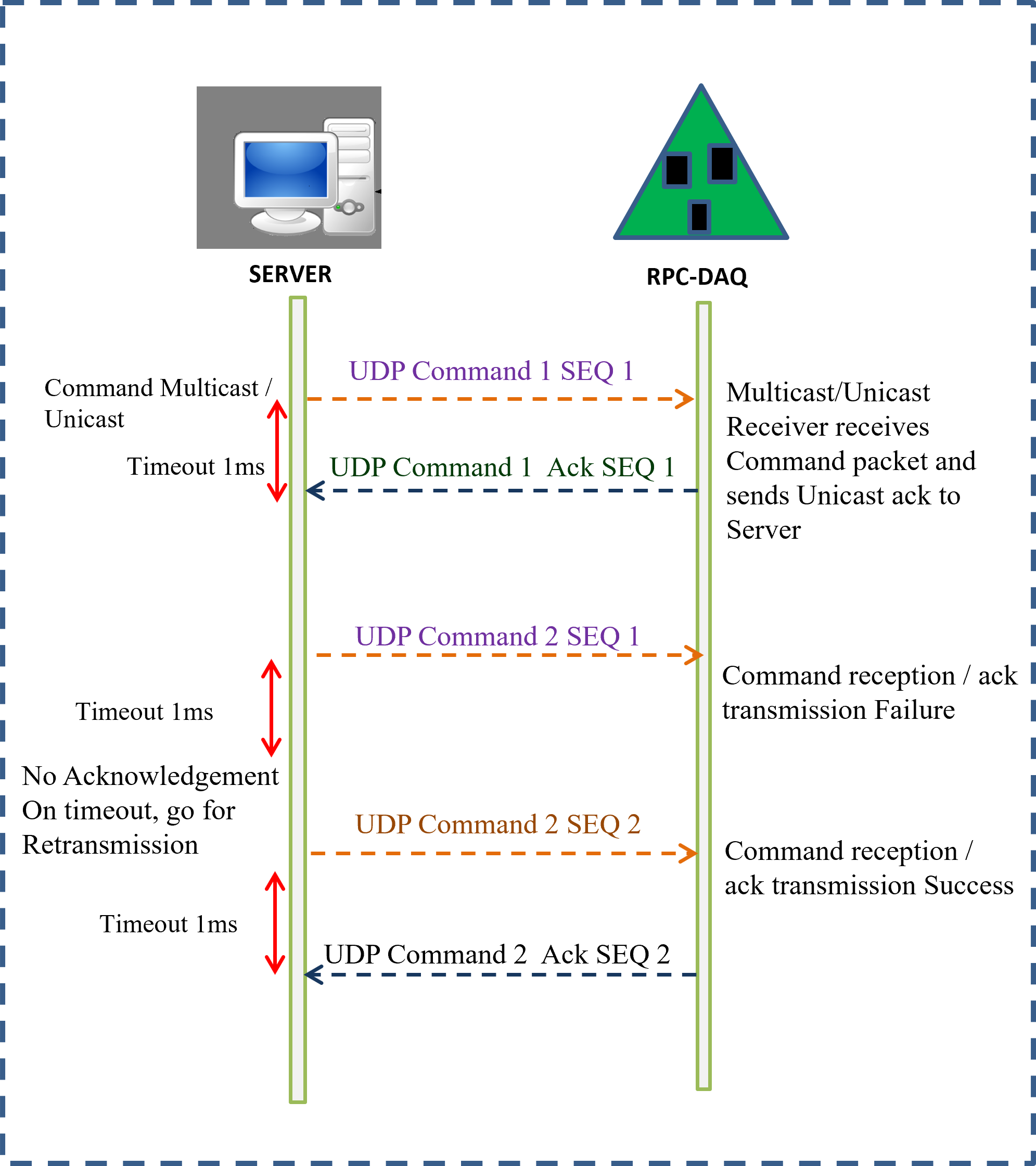} 
\caption{Command Data Flow between Back-end and RPC-DAQ.}
\label{fig:2a}
\end{subfigure} 
\quad
\begin{subfigure}[c]{0.4\textwidth}
\includegraphics[width=\linewidth]{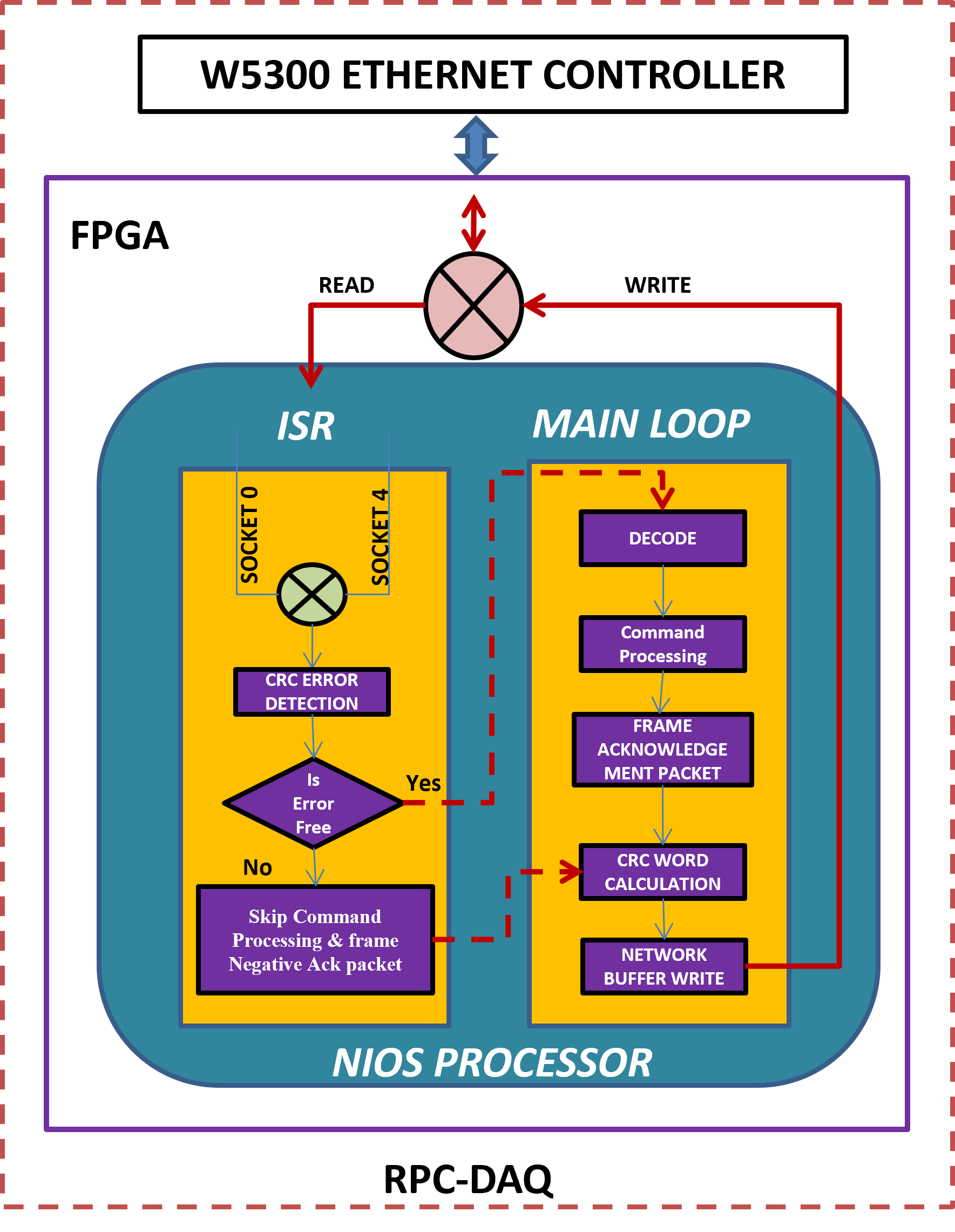}
\caption{UDP ISR and Main loop handshake.}
\label{fig:2b}
\end{subfigure}
\caption{Command Processing in the Server and NIOS Processor.\label{fig:2}}
\end{figure}

The RPC-DAQ's control and configuration are managed through remote commands from the back-end console. These commands are transmitted as packets via UDP command interface sockets. The RPC-DAQ processes two types of commands namely group commands and DAQ specific commands. The group commands are issued to all or group of DAQ nodes using UDP Multicast whereas DAQ specific commands are used to control individual DAQ nodes using UDP Unicast. Each RPC-DAQ unit initializes UDP servers during startup. Upon receiving a command from the back-end server, the processor executes a specific task on the RPC-DAQ, and acknowledges the operation. Dedicated handshaking and checksum schemes, as depicted in Figure~\ref{fig:2}(a), are implemented to maintain the reliability of commands.

The INO-ICAL experiment~\cite{d} is composed of three modules, each comprising of 9,600 detector units. These units are grouped and routed to six central command and monitoring servers via two tiers of Gigabit Ethernet switches. Each detector unit and its corresponding RPC-DAQ will have its dedicated database managed by its respective command server. To support the event building, RPC-DAQs will be grouped according to their geometric location with a common Multicast address. They use a UDP Multicast socket to receive common group commands from the server. Each DAQ opens this socket in Multicast mode by registering to a common Multicast group IP address, such as 239.0.0.1, through IGMP/ICMP. Additionally, a Unicast socket receiver and transmitter facilitate the reception of DAQ specific commands and the transfer of acknowledgments. Using these two sockets, back-end servers can communicate with groups of DAQs or individual DAQs using the respective group IP or individual IP addresses.

\begin{figure}[htbp]
\centering
\includegraphics[width=.8\textwidth]{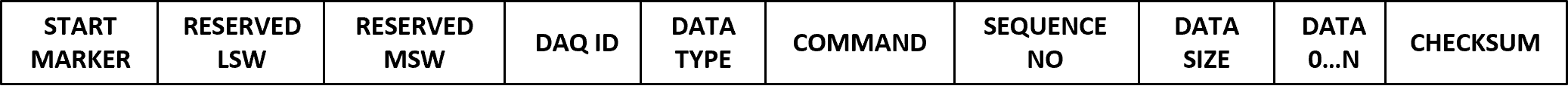}
\caption{Command Format Between Front-end and Back-end.\label{fig:3}}
\end{figure}

Common group commands, such as "ISDAQUP", "LOADRTC", and "ENABLEEVENT", are used primarily for run control, changing data acquisition settings, configuring network parameters, and monitoring parameters. Depending on the command type,  payload data and the UDP packet size vary from 18 to 100 bytes. To improve reliability, a 16-bit CRC checksum word is added to the payload of both the command packet sent from the server and the acknowledgment packet from the RPC-DAQ. Figure~\ref{fig:3} shows a typical UDP command packet.  

Upon receiving a command packet, the Ethernet controller in the RPC-DAQ invokes an ISR for command execution, as illustrated in Figure~\ref{fig:2}(b). The worst-case command processing time is 615 microseconds when the NIOS processor runs at a clock speed of 50 MHz. Event and monitoring data transfer between each RPC-DAQ and the back-end server~\cite{e} are facilitated through TCP/IP client-server connections, initiated using the UDP command interface. When a remote Firmware upgrade is required, a back-end UDP command is used to establish TCP/IP server-client connection. The command interface primarily serves for control and monitoring of detector parameters like high voltage settings, current monitoring, detector configurations and network configurations. Users can modify network settings like IP addresses, Multicast addresses, port numbers, and socket modes (TCP/UDP) of any DAQ unit with proper authentication and permissions. Checksum based error detection and retransmission during negative acknowledgments ensures this command protocol is more reliable than only using the UDP, and thus it is named as Hybrid Protocol-based Command Interface (HPCI).

\section{Software Controlled Event Data Acquisition Scheme}
\label{sec:conf}

\begin{figure}[htbp]
\centering
\includegraphics[width=.8\textwidth]{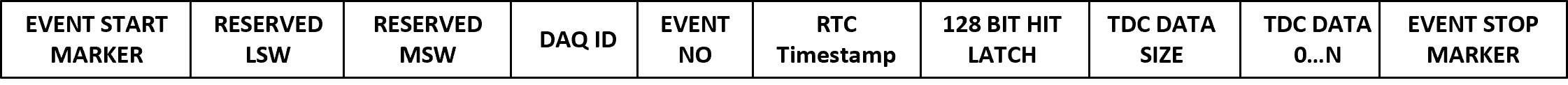}
\caption{Event Packet Format.\label{fig:4}}
\end{figure}

\begin{figure}[htbp]
\centering
\begin{subfigure}[c]{0.5\textwidth}
\includegraphics[width=\linewidth]{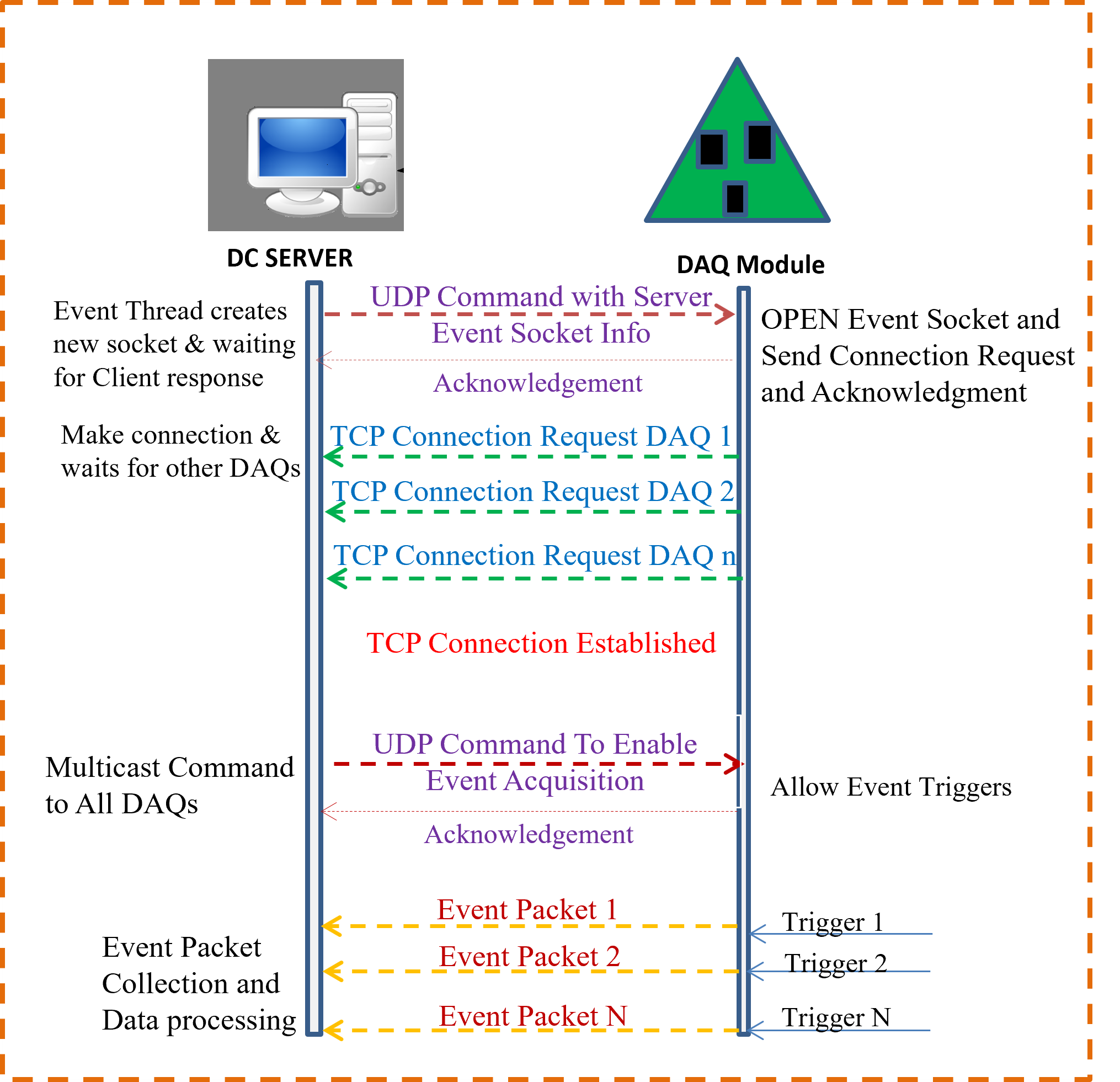} 
\caption{Event Data transfer between Front-end and Back-end.}
\label{fig:5a}
\end{subfigure}\hfill  
\begin{subfigure}[c]{0.45\textwidth}
\includegraphics[width=\linewidth]{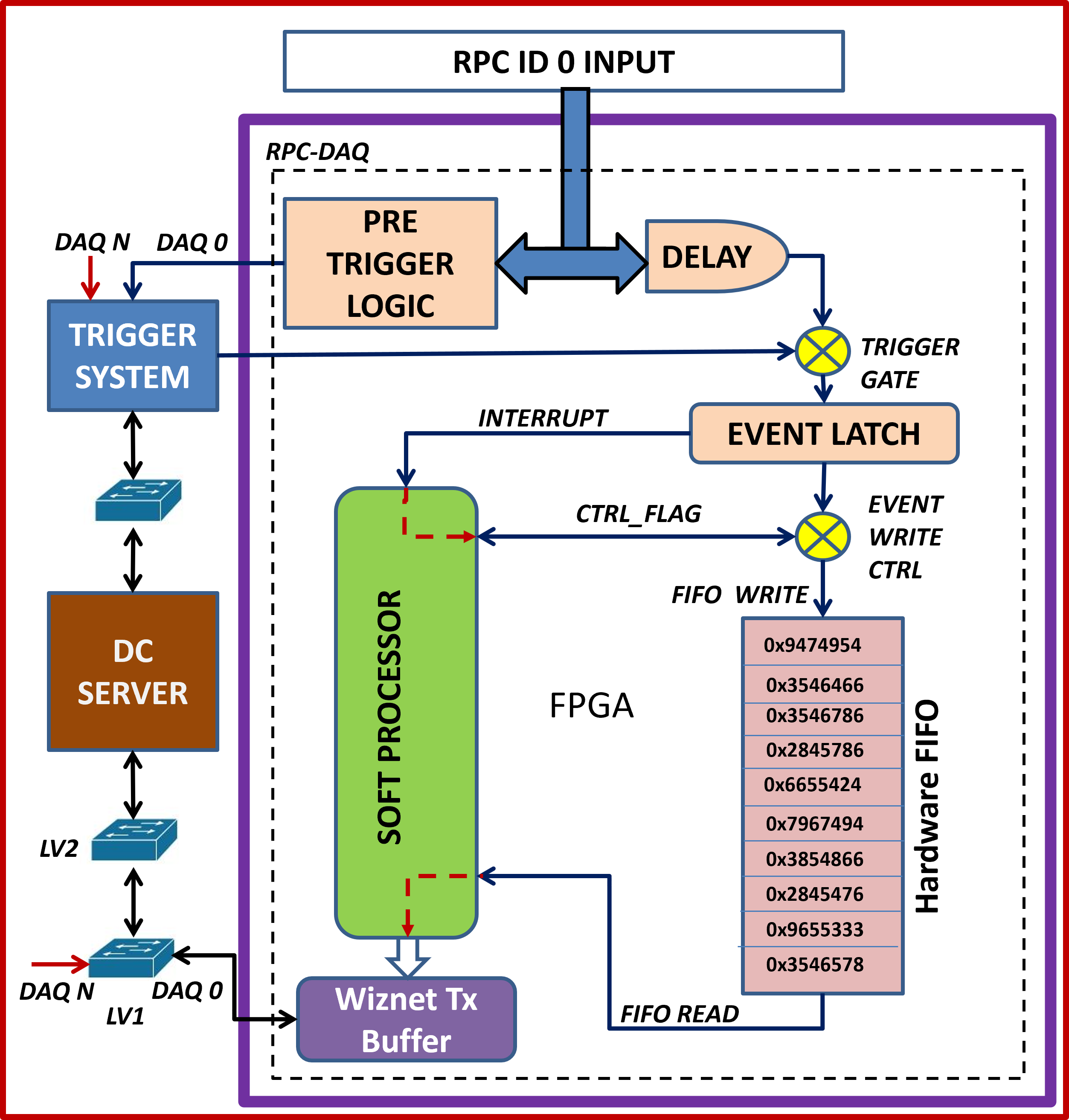}
\caption{SCEDA Overview.}
\label{fig:5b}
\end{subfigure}
\caption{Event Data Acquisition using SCEDA. \label{fig:5}}
\end{figure}

Each RPC-DAQ unit establishes a TCP/IP connection with the Data Concentrator (DC)~\cite{f}, as depicted in Figure~\ref{fig:5}(a). Currently there is only one Data Concentrator to handle events from 20 RPC-DAQs.  At the start of the run a UNICAST command "SETEVETCPSOCK" is sent to every RPC-DAQ sequentially to initiate this connection process. Once the connection is established on every event trigger transmitted to the RPC-DAQs, the NIOS processor software receives an event interrupt.  The Event ISR function packs the event data of varying lengths and writes it into an hardware FIFO. The data from the event FIFO is transferred to the Ethernet controller's transmit buffer in the main loop as a packet of optimal size. The Ethernet controller then transmits the event data to the back-end DC over the LAN . The Event data packet shown in Figure~\ref{fig:4} are marked with event numbers using a global Real Time Clock (RTC) timestamp at DC. Every RPC-DAQ runs an RTC using a global clock and has a precision of $100\, \text{ns}$. The RTC of each DAQ is synchronized with the a global RTC at the start of each run. These timestamps are latched along with event data during trigger processing.

To handle the high trigger rate, the event dead-time was minimized by implementing a Software-Controlled Event Data Acquisition (SCEDA) scheme, as shown in Figure~\ref{fig:5}(b). When an event interrupt is received, the registered event data is written into the hardware event FIFO within the event ISR. Depending on the remaining space in the event FIFO the decision will be taken to allow or block the incoming triggers. The data from the event FIFO is transferred to the Ethernet transmit buffer under software control in the background, and a transmit command to the Ethernet controller is issued to send the data over the Ethernet link. If the trigger isn't enabled in the event ISR due to a lack of empty space in the FIFO for the next trigger, the trigger is enabled after sufficient space is created by transferring the event data to the network buffer. With this scheme, as shown in Table~\ref{tab:2}, an event dead-time of $350\, \text{ns}$ is achieved for the maximum event data size of 600 bytes. 

\begin{table}[htbp]
\centering
\caption{Deadtime calculated for acquiring 600 Bytes of event data.\label{tab:2}}
\smallskip
\begin{tabular}{l|c}
\hline
Process &Timetaken (Microsecond)\\
\hline
Time taken for reading event data and writing into hardware FIFO  & 31 \\
Transferring event data from HW FIFO to Ethernet FIFO & 190\\
Packet Transfer over Ethernet & 60\\
\hline
Total Dead Time & 281\\
\hline
\end{tabular}
\end{table}
Global physics triggers from the ICAL detector occur randomly with an average frequency of 2 kHz. The event data size ranges from 80 to 600 bytes, depending on the number of hits in the detector, which in turn is influenced by the detector's multiplicity and noise levels. The average event size is about 120 bytes. The packet data transfer over the LAN network usually takes around 60 microseconds depending on network traffic and congestion. If there's no network traffic, data transfer over the network occurs concurrently with event data acquisition into the network buffer. 
            % &281 ($\mu$s) & 125 ($\mu$s) \\
\begin{table}[htbp]
\centering
\caption{Event Loss Percentage due to dead time with Varying data size. \label{tab:3}}
\smallskip
\begin{tabular}{l|c|c}
\hline
Event Size & Dead Time & Dead Time\\
            &281 ($\mu$s) & 125 ($\mu$s) \\
\hline
120 Bytes & 0.0006 & 0\\
408 Bytes & 0.0437 & 0\\
600 Bytes & 24.8 & 0.0014\\
Random Event Size(80 to 600 Bytes) & 0.0008 &0.0022\\
\hline
\end{tabular}
\end{table}

Table~\ref{tab:3} illustrates the trigger losses resulting from event dead-time for various event data sizes at a random trigger rate of 5 kHz. To improve the trigger rate and minimize loss percentages it is essential to reduce the event data transfer time from the hardware FIFO to the Ethernet FIFO. According to the Wiznet W5300 data sheet~\cite{g}, the write cycle time for 16bit data is approximately $80\, \text{ns}$. But due to the control signals generated by the NIOS software the write cycle time goes up to $350\, \text{ns}$. This can be improved by reducing Wiznet write cycle time. The Wiznet FIFO is accessed using an handshake logic which shifts Wiznet write access to hardware from the processor only when necessary. This decreases the FIFO data transfer time from $190\, \text{$\mu$s}$ to $35\, \text{$\mu$s}$. As a result the overall dead time was reduced from $281\, \text{$\mu$s}$ to $125\, \text{$\mu$s}$. This scheme has been tested only in the test-bench and has not been deployed in the real experiment yet.

\section{Health Monitoring}
\label{sec:flow}
Periodic monitoring of RPC health and RPC-DAQ ambient parameters are important. This is carried out through monitoring of key parameters like RPC strip noise rates, fold rates, high voltage, currents and ambient conditions such as temperature, pressure, and humidity. The DAQ logic uses a programmable monitoring timer that generates monitoring interrupts, holding an interrupt priority of 3 following the event interrupt and command UDP interrupt. Typically, the timer is set to 10 seconds during event runs. The lower monitor trigger frequency helps in reducing the load on event dead time because monitoring process usually takes few hundreds of milliseconds.

The Monitoring ISR reads 16 strip count rates and 9 fold count rates from DAQ logic scalers each of length 32 bits with first 8 bits marked as channel ID. This data along with on-board temperature, pressure and humidity sensor data are packed into a software array and then transferred to the Mon Socket transmit buffer.

\begin{figure}[htbp]
\centering
\includegraphics[width=.7\textwidth]{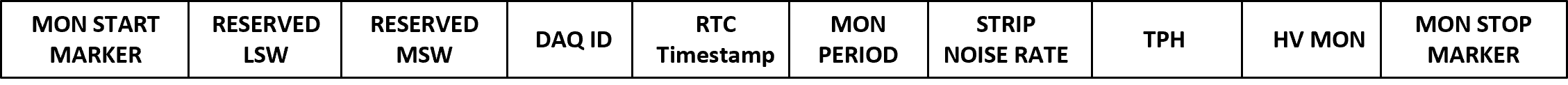}
\caption{Monitoring Packet Format.\label{fig:6}}
\end{figure}
The monitoring data packets shown in Figure~\ref{fig:6} are transmitted to the back-end server through a dedicated TCP/IP socket, which is established at the beginning of the run.
\begin{figure}[htbp]
\centering
\includegraphics[width=.7\textwidth]{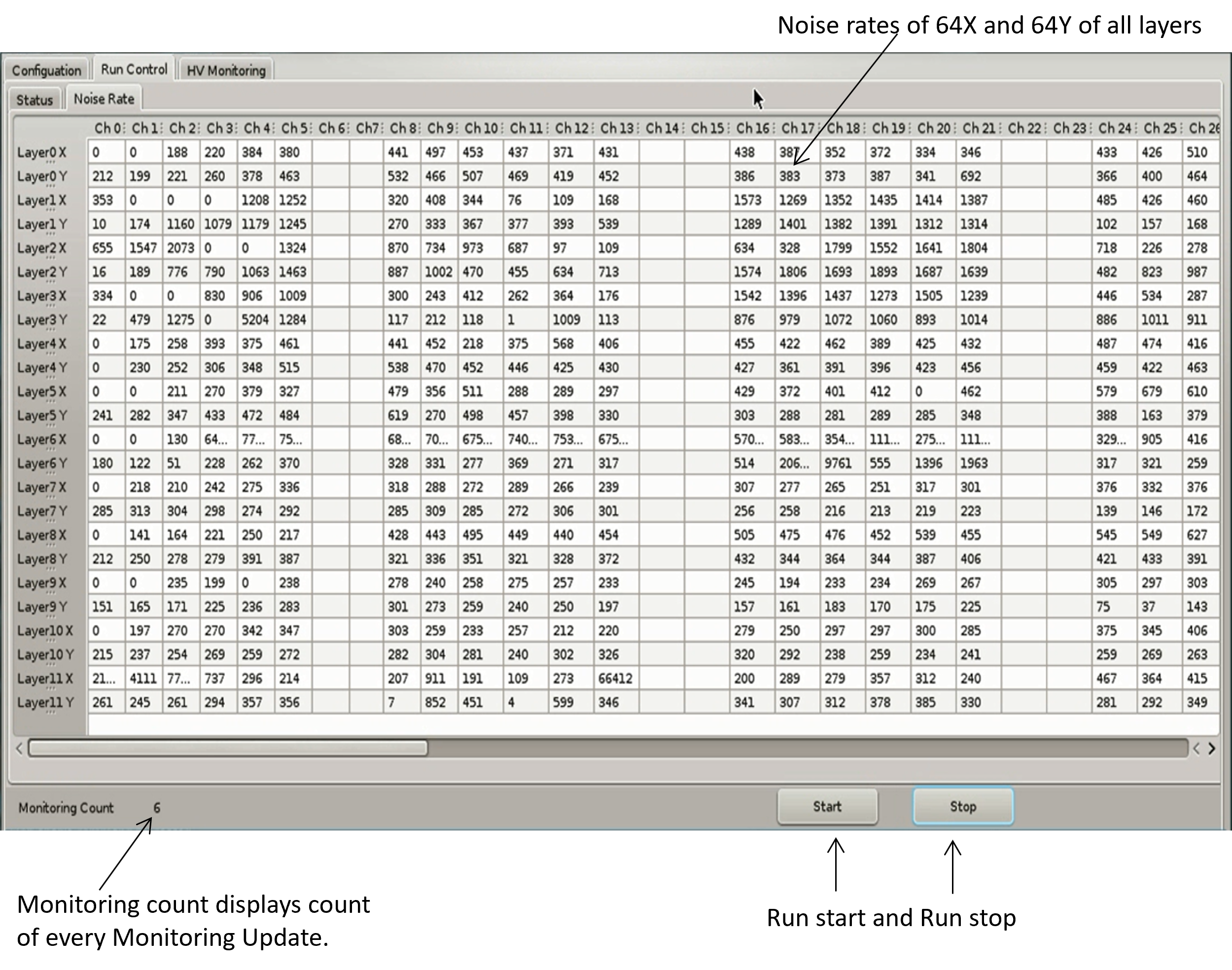}
\caption{Rate Monitoring Software.\label{fig:7}}
\end{figure}
In the back-end monitoring software shown in Figure~\ref{fig:7} is used to display and store the monitoring data collected from all RPCs. Users have the access to monitor the detector's health in real-time. Rate display are used to identify the channels which has high noise rates. Also strip masking commands are used to deactivate the noisy channel in the respective DAQ.  A dedicated "MONITORING RUN" feature is available to the user for debugging the detector during installation. The accumulated monitoring data can be scrutinized to evaluate the overall health of the RPCs, with any irregularities or faults being documented and flagged for further attention.

\section{HV Control and Monitoring Software}
\label{sec:Result}
\begin{figure}[htbp]
\centering
\includegraphics[width=.8\textwidth]{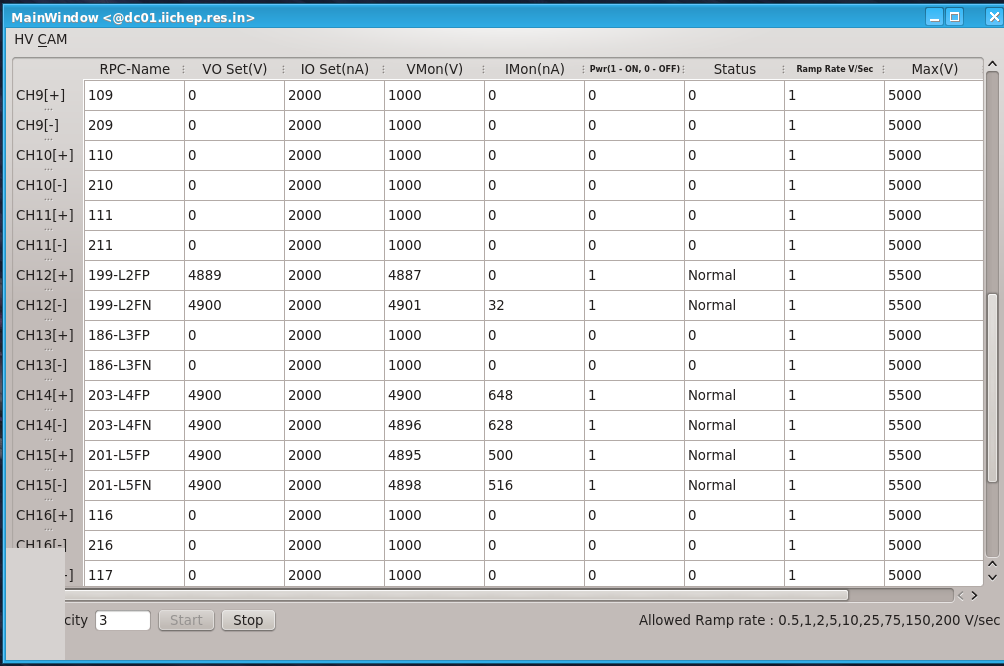}
\caption{High Voltage Control and Monitoring software.\label{fig:8}}
\end{figure}

A specialized INO HV module capable of providing a voltage range from 0 to $\pm{6}$ kV, is positioned in close proximity to the detector. It is linked to the  RPC-DAQ via a standard SPI protocol. To facilitate remote communication with the HV module a dedicated HV control and monitoring software shown in Figure~\ref{fig:8} was developed using PyQt4. This software employs the HPCI command architecture to issue HV control related UDP commands to the RPC-DAQ.

Users can configure the high voltage of desired RPC through the graphical user interface (GUI), which subsequently converts the values into UDP packets for transmission to the corresponding detectors. Upon receiving HV commands such as "HV ON", "OVERVOLTAGE SET", "VOLTAGE SET" and "HV MON" from the command server, the RPC-DAQ NIOS processor dispatches SPI commands to the HV module. Each SPI command takes around 80 milliseconds to make changes in the HV module settings. Care must be taken to avoid HV commands during event runs to minimize dead time impact. The software efficiently manages HV of multiple RPCs simultaneously using UDP UNICAST protocol.

The GUI incorporates a slow monitoring algorithm that displays the observed values. The monitored high voltage (HV) is logged in a log file along with timestamps. During the HV ramping up and ramping down phases the software autonomously identifies any over current and adjusts the voltage accordingly. Currently, a small-scale version of the Control and Monitoring Server has been developed for testing the functionality of the high voltage sub-system using NIOS for 20 RPCs(40 Channels). In the INO-ICAL experiment, each of the three detector modules will consist of 9600 RPC units. A Central Control and Monitoring Server will manage these units in segmented configurations. Each server will maintain a dedicated detector database for its assigned segment. To access a specific detector unit, user will first connect to the server overseeing the corresponding detector segment.

\section{Remote Firmware Upgradation}
\label{sec:rfu}
In mega-science projects like ICAL, which has a network of 28,800 RPC-DAQs distributed across a large area, the traditional method of upgrading FPGA firmware using JTAG interfaces and lab programmer kits becomes unfeasible. An innovative firmware upgrade process is implemented to configure remotely located RPC-DAQs. This process uses an TCP/IP based client-server socket connection within the existing network to upload new firmware to the DAQs. 

\begin{figure}[htbp]
\centering
\includegraphics[width=.5\textwidth]{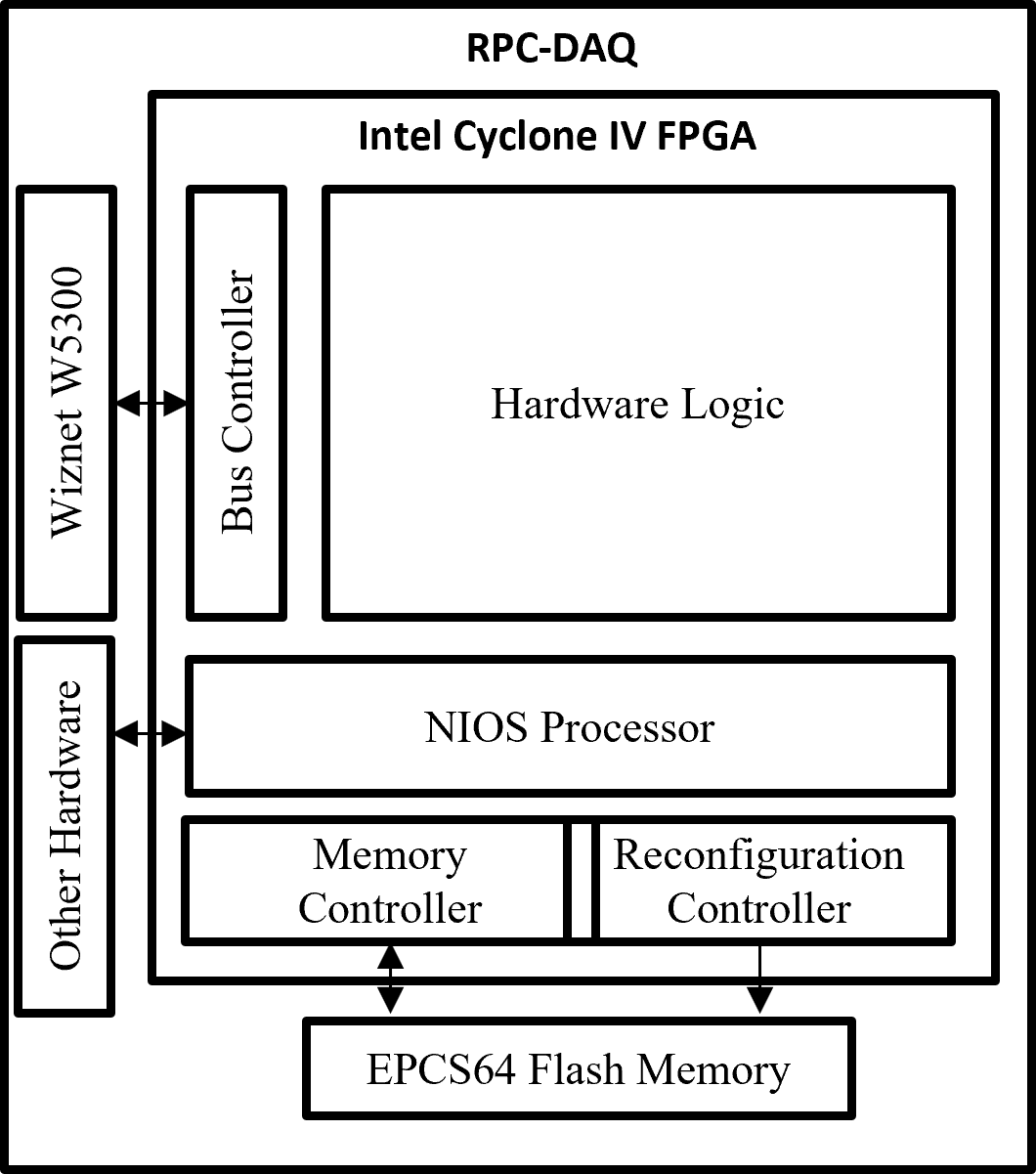}
\caption{RPC-DAQ Block Diagram Highlighting Remote Firmware Upgradation.\label{fig:9}}
\end{figure}

A block diagram depicting the various components in RPC-DAQ used in firmware upgradation process is shown in Figure~\ref{fig:9}. The core of this process consists of writing the new FPGA firmware to a distinct page of the flash memory (EPCS64)~\cite{h} within each DAQ node. The page is then configured for the next reboot. A remote command can also be used to initiate a forced reboot which deploys the new firmware. This Remote Firmware Upgradation (RFU) procedure is resource efficient and less time consuming.

To maintain the reliability of firmware transfer a specialized handshake protocol on top of the TCP/IP is employed. RFU uses four distinct packets as outlined in Table~\ref{tab:4}. These packets are retransmitted to the node in case of any packet loss by consistently monitoring of the packet sequence number. Upon receiving the "Begin" packet the processor erases the corresponding image location in the flash memory based on the decoded packet information. An acknowledgment is then sent to the back-end to commence the transfer of firmware data using "Data" packet. Once the entire firmware data is transferred to the flash the back-end sends a "Fin" (finish) packet to the DAQ signifying the completion of the firmware data transfer. Following this the DAQ closes the RFU socket and resumes its routine main loop.

\begin{table}[htbp]
\centering
\caption{Packet Datatypes used in Remote Firmware Upgradation.\label{tab:4}}
\smallskip
\begin{tabular}{l|c|l}
\hline
Packet Type & Size(Bytes) & Payload Description\\
\hline
Begin Packet & 10 & Firmware file size and image number\\
Data Packet & 1280 & Firmware Data Transfer\\
Fin Packet & 4 & Indicator\\
Ack Packet & 8 & Status\\
\hline
\end{tabular}
\end{table}
Data packet reliability is ensured through XOR checksum verification. RFU allows users to configure the flash memory with multiple firmware images offering flexibility for recovery and reconfiguration. Specialized UDP commands facilitate reconfiguration and image selection. The RFU scheme is explained in detail in ~\cite{i}. Configuring a single RPC-DAQ module takes approximately 21 seconds. With the current RFU scheme and multithreading, up to 10 RPC-DAQ modules can be configured simultaneously in roughly the same 21-second time-frame. This can be scaled to 100 modules by processing smaller groups of DAQs in batches, with each batch being configured sequentially.

% 21 seconds required to program 1 DAQ
% Using 1 server one can program 10 DAQs simultaneously. 

\section{NIOS Software Testing and Commissioning in Mini-ICAL}
\label{sec:mical}
\begin{figure}[htbp]
\centering
\includegraphics[width=.8\textwidth]{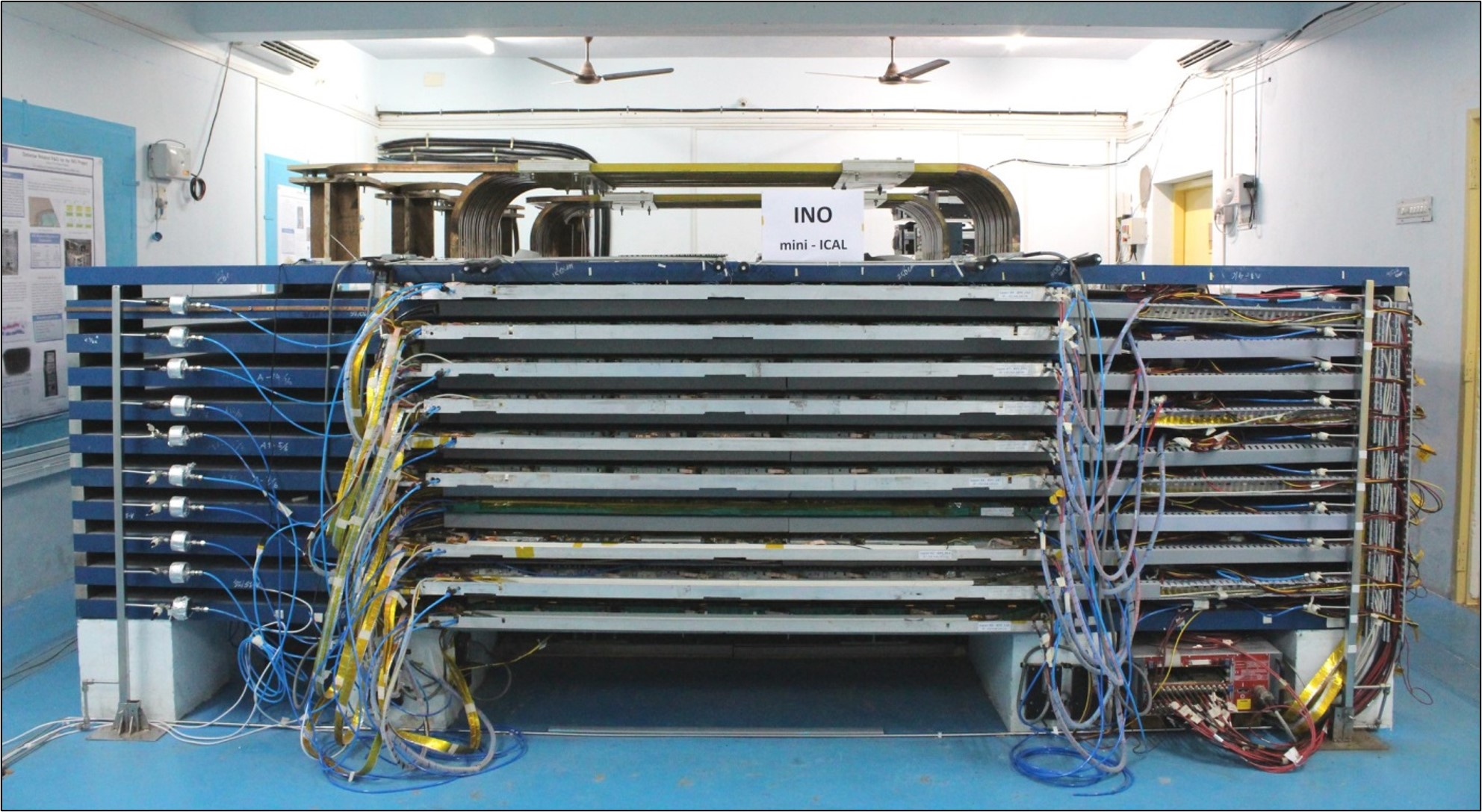}
\caption{Mini-ICAL Experiment with RPC Detector Units.\label{fig:10}}
\end{figure}

Following the successful production of RPC-DAQ modules, the master firmware of the RPC-DAQ which consists of the NIOS softcore is set as the factory image of all the DAQs. These RPC-DAQs are integrated into Mini-ICAL (mICAL)~\cite{j} shown in Figure~\ref{fig:10}, a prototype for the proposed ICAL consisting of stacked iron plates to accommodate the RPCs and Front-end electronics~\cite{k}. The scope of mini-ICAL includes studying and characterizing detectors and readout electronics under the designed magnetic field of approximately 1.3 Tesla for the INO-ICAL detector. It also aims to address engineering challenges such as cable harnessing, routing, spatial constraints, grounding, shielding and noise mitigation. While the industry produced electronics boards and modules are bench-tested using specially developed test jigs, they will be also batch-tested and characterized under experimental conditions using mini-ICAL detector. So far 25 RPC-DAQ modules are fabricated and 20 RPC-DAQs are being used in the mini-ICAL experiment. Initial testing of the RPC-DAQs focuses on health monitoring, evaluating the functionality of the NIOS, Wiznet, and software components. The response to various commands is verified, with acknowledgment packets using HPCI.

Before installation every RPC-DAQ module undergoes testing using a specially designed test jig module for all functionalities. Event wise realistic data patterns are generated using these test jig modules and fed to the RPC-DAQs to assess data integrity. During mICAL commissioning, an in-situ trigger generation setup was prepared to test the event handling capability of the RPC-DAQ in a real system. A fixed frequency trigger system is initially employed to trigger the RPC-DAQs for event acquisition and the resulting data are verified.

In the mICAL version of the firmware, the clock speed is maintained at 50 MHz and the overall event acquisition process incurs a dead time of around 250 to 500 microseconds. The trigger system is configured with a 500 microsecond trigger blocking to avoid partial event loss. The HV control and monitoring software was successfully deployed and tested to verify the functionality of NIOS software in controlling the HV of the RPCs. Realistic testing of remote firmware upgrading was conducted in mICAL where the RPC-DAQs are placed inside the iron stack with only Ethernet access. The RFU software at the back-end and NIOS processor efficiently configured the inaccessible RPC-DAQs. This software allows simultaneous configuration up to 10 RPC-DAQs. The version of the RPC-DAQ reported in this paper is finalized and ready for production. Large-scale production will commence as soon as the final project clearance is obtained. 

%Scope:
% The Scope of MICal is to study and characterize detector and readout electronics under magnetic filed of around 1.3 Tesla. it is also used to Study the engineering challenges like Cable routing, space, grounding and noise. 
% line 203: So far 25 RPC-DAQ modules are fabricated
% line 223: Yes , this was the plan for production and by increasing CPU core/Threads  and servers one can achieve  more number of DAQs

\section{Other Responsibilities of NIOS in ICAL Experiment}
\label{sec:other}
\paragraph{Initialization of Network Configuration Data}
In RPC-DAQ the flash memory is used during the boot sequence for reading and writing network configuration data. The Factory firmware allows users to write predefined network configuration data such as IP address, Gateway, and MAC to the flash memory. These details are read during boot-up to configure the Wiznet ethernet controller. Users can send the "NETWORK CONFIG" Command at any time to modify the network settings of the RPC-DAQ from the command server.

\paragraph{HPTDC Configuration}
A primary function of the RPC-DAQ is to accurately timestamp event hits using the High Performance Time to Digital Converter (HPTDC) ASIC~\cite{l}. Control and configuration of the HPTDC are executed through the NIOS Processor. During the boot sequence, HPTDC configuration data is stored in a register array in the NIOS processor. This data is then read and configured to the HPTDC using a JTAG interface. Additionally, the "HPTDC CONFIG" Command from the back-end can remotely configure new HPTDC configurations in the RPC-DAQ.

\paragraph{DAQ Logic Configuration}
The RPC-DAQ logic registers are accessed through a 32bit Logic bridge controller. The NIOS processor uses this controller to access the RTC timestamp, event hit information and FIFO.

\section{Conclusion}
RPC-DAQs are used in mICAL supporting daily event data acquisitions. The NIOS processor along with its versatile features facilitates a range of functionalities to reduce FPGA hardware logic complexities. Soft-core processors like NIOS prove highly suitable for both standalone and distributed data acquisition in extensive experiments like ICAL. The NIOS achieves an optimal trigger rate of 5 kHz with a typical event size of 120 Bytes. Further reduction in event dead-time down to $125\, \text{$\mu$s}$ is achievable through hardware access of Ethernet controller Wiznet W5300. The current design of the system is found to meet the trigger and data throughput requirements of the INO-ICAL experiment as discussed in Section 1 and has undergone successful testing within the mICAL. 
% line 245 : typical event size:
%120 bytes

% line 247 : Requirements:
% Intro
% The RPC-DAQ in the INO ICAL experiment must handle a minimum trigger rate of 2 kHz and transmit event packets containing position and timing information at a throughput of up to 2 Mbps, with a typical event size of 120 bytes. Additionally, it is crucial to maintain simultaneous slow control and health monitoring while acquiring event data. 

\acknowledgments
We sincerely thank all our present and former INO colleagues especially Puneet Kanwar Kaur, Umesh L, Upendra Gokhale, Anand Lokapure, Aditya Deodhar, Suraj Kole, Rajkumar Bharathi for technical support during the design, Also we would like to thank members of TIFR namely Pavan Kumar, S.R. Joshi, Piyush Verma, Darshana Gonji, Santosh Chavan and Vishal Asgolkar who supported testing and commissioning. Also we like to thank Former INO directors N.K. Mondal and V. M Datar for their continuous encouragement and guidance.

\end{document}